\documentclass[fleqn,usenatbib]{mnras}
\usepackage{newtxtext,newtxmath}

\usepackage[T1]{fontenc}

\usepackage{graphicx,natbib,gensymb,eurosym,wrapfig,layout,paralist,tabularx,booktabs,enumitem,setspace,multicol,hologo,textcomp,amsmath,hyperref,url,mathtools,subfig,float,amssymb,xspace,comment,lastpage,color,soul}

\title[Mode changing in J1909$-$3744]{Mode changing in  J1909$-$3744: the most precisely timed pulsar}

\author[M.~T.~Miles et al.]{
M.~T.~Miles,$^{1,2}$\thanks{E-mail: \href{mailto:matthewmiles@swin.edu.au}{matthewmiles@swin.edu.au}}
R.~M.~Shannon,$^{1,2}$
M.~Bailes,$^{1,2}$
D.~J.~Reardon,$^{1,2}$
S.~Buchner,$^{3}$\newauthor
H.~Middleton,$^{1,2,4}$
R.~Spiewak,$^{1,2,5}$
\\
$^{1}$Centre for Astrophysics and Supercomputing, Swinburne University of Technology, PO Box 218, Hawthorn, VIC 3122, Australia\\
$^{2}$ARC Centre of Excellence for Gravitational Wave Discovery (OzGrav), Mail H29, Swinburne University of Technology, PO Box \\ 218, Hawthorn, VIC 3122, Australia \\
$^{3}$South African Radio Astronomy Observatory, 2 Fir Street, Black River Park, Observatory 7925, South Africa\\
$^{4}$School of Physics, University of Melbourne, Parkville, VIC 3010, Australia\\
$^{5}$Jodrell Bank Centre for Astrophysics, Department of Physics and Astronomy, University of Manchester, Manchester M13 9PL, UK
}

\date{Accepted XXX. Received YYY; in original form ZZZ}

\pubyear{2021}

\begin{document}
\label{firstpage}
\pagerange{\pageref{firstpage}--\pageref{lastpage}}
\maketitle

\begin{abstract}
    We present baseband radio observations of the millisecond pulsar J1909$-$3744, the most precisely timed pulsar, using the MeerKAT telescope as part of the MeerTime pulsar timing array campaign. During a particularly bright scintillation event the pulsar showed strong evidence of pulse mode changing, among the first millisecond pulsars and the shortest duty cycle millisecond pulsar to do so. Two modes appear to be present, with the weak (lower signal-to-noise ratio) mode arriving $9.26$ $\pm 3.94$ $\mu$s earlier than the strong counterpart. Further, we present a new value of the jitter noise for this pulsar of $8.20 \pm 0.14$ ns in one hour, finding it to be consistent with previous measurements taken with the MeerKAT ($9 \pm 3$ ns) and Parkes ($8.6 \pm 0.8$ ns) telescopes, but inconsistent with the previously most precise measurement taken with the Green Bank telescope ($14 \pm 0.5$ ns). Timing analysis on the individual modes is carried out for this pulsar, and we find an approximate $10\%$ improvement in the timing precision is achievable through timing the strong mode only as opposed to the full sample of pulses. By forming a model of the average pulse from templates of the two modes, we time them simultaneously and demonstrate that this timing improvement can also be achieved in regular timing observations. We discuss the impact an improvement of this degree on this pulsar would have on searches for the stochastic gravitational wave background, as well as the impact of a similar improvement on all MeerTime PTA pulsars.
\end{abstract}


\section{Introduction}
Observing and monitoring the emission of millisecond pulsars (MSPs) has enabled novel tests of general relativity \citep{1982ApJ...253..908T, 2003Natur.426..531B, Kramer_double_pulsar_2006}, led to the discovery of planetary systems orbiting distant stars \citep{Wolszczan_Frail_1992, Wolszczan_planets_1994} and put constraints on the nuclear equations of state \citep{Demorest_Nature_2010, Antoniadis1233232, 2021ApJ...915L..12F, 2021arXiv210506979M, 2021arXiv210506980R}. In the coming years, this practice will also likely detect nHz frequency gravitational radiation \citep{Hellings_Downs_1983}, with recent publications showing potential evidence of common spectral processes present 
in timing data
\cite[][]{NanoGravGWB, 2021arXiv210712112G}. 

The vast majority of these discoveries have come from the process of precision pulsar timing \citep{bailes_2009}, in which the observed pulsar emission is cross-correlated with a representative template, and the arrival times and comparative uncertainties are modelled. This process usually assumes that the pulse profiles at a given frequency are identical at each epoch, and that the only source of noise is radiometer noise associated with the non-zero temperature of the receiving system and the sky. However, this is not necessarily the case as pulsar emission is known to vary on many timescales. Variations can include giant pulses that exhibit a far greater flux than would ordinarily be expected of the pulsar \citep{Lundgren_1995_crab_giant, Cognard_1997_giant_J1937, Geyer_TPA_2021_J0540}, abrupt changes in the observed emission profiles \citep{Shannon_2016_profilechange}, and sub-pulse drifting \citep{2006A&A...445..243W}. 
Pulse morphology can also be influenced by factors that are external to the pulsar, such as the changing interstellar medium between the Earth and the source that can lead to effects such as pulse broadening \citep{2010arXiv1010.3785C}, or lensing from surrounding material \citep{2001MNRAS.321...67L}.

There are also the phenomena of mode changing and nulling \citep{1982ApJ...258..776B, Wang_2007_nulling_changing}, where the emission strength and morphology of the profile changes in sometimes periodic and sometimes irregular cadences, or even ceases completely \citep{1970Natur.228.1297B, 2010MNRAS.408L..41T}. These events have previously been linked to spin-down events in slow pulsars thought to be related to changes in magnetospheric currents \citep{2006Sci...312..549K,2010Sci...329..408L}, causing some pulsars to change their spin down by up to $50\%$ when changing between modes. To date, only two millisecond pulsars, PSR B1957$+$20 and PSR J0621$+$1002, have shown evidence of mode changing \citep{2018ApJ...867L...2M, 2021ApJ...913...67W}. MSPs are generally faint, making it difficult to study individual pulses, so it is currently unclear whether these pulsars represent a wider population or are unique. Profile variations stemming from mode changing will affect pulse arrival time measurements, and lead to a decrease in timing precision and potentially bias the measurement of arrival times (ToAs) \citep{2014MNRAS.443.1463S, 2017MNRAS.466.3706L}.

In addition to showing discrete changes in shape, it has been known since the first observations of pulsars that individual pulses show stochastic pulse variations from pulse to pulse. This is manifested in the arrival times as pulse jitter noise, which is an excess white noise term in the pulse arrival times \citep{2014MNRAS.443.1463S, 2019ApJ...872..193L, 2021MNRAS.502..407P}. Accounting for jitter noise is crucial to the science of pulsar timing and the search for the stochastic gravitational wave background (SGWB), as it implies that more sensitive observations do not necessarily provide greater timing precision. Ultimately, this limits the timing precision that can be achieved at high gain facilities, and impacts the sensitivity of pulsar timing arrays (PTAs) \citep{1978SvA....22...36S, 1979ApJ...234.1100D, 1990ApJ...361..300F}.

In this work we examine the single pulse variability of the MSP J1909$-$3744 using observations taken with the MeerKAT Array. We show that this MSP is the third of its kind to show strong evidence of mode changing. Additionally, we show that there is a modest improvement in timing precision by only including the bright pulses. In Section \ref{sec: Observations} we describe the data and the MeerTime observing systems in brief. In Section \ref{sec: mode_changing} we analyse the emission structure and present evidence for bimodality in the pulse emission brightness and pulse emission state, discussing its impact on timing precision. In Section \ref{sec: Jitter Variation} we estimate the jitter noise and compare it to previously published values. In Section \ref{sec: Discussion} we discuss our results, and briefly explore the impact of changing the standard pulsar timing methodology to a baseband timing method where possible, demonstrating the feasible PTA sensitivity improvements achievable through this.

\begin{figure}
    \centering
    \includegraphics[width=\columnwidth]{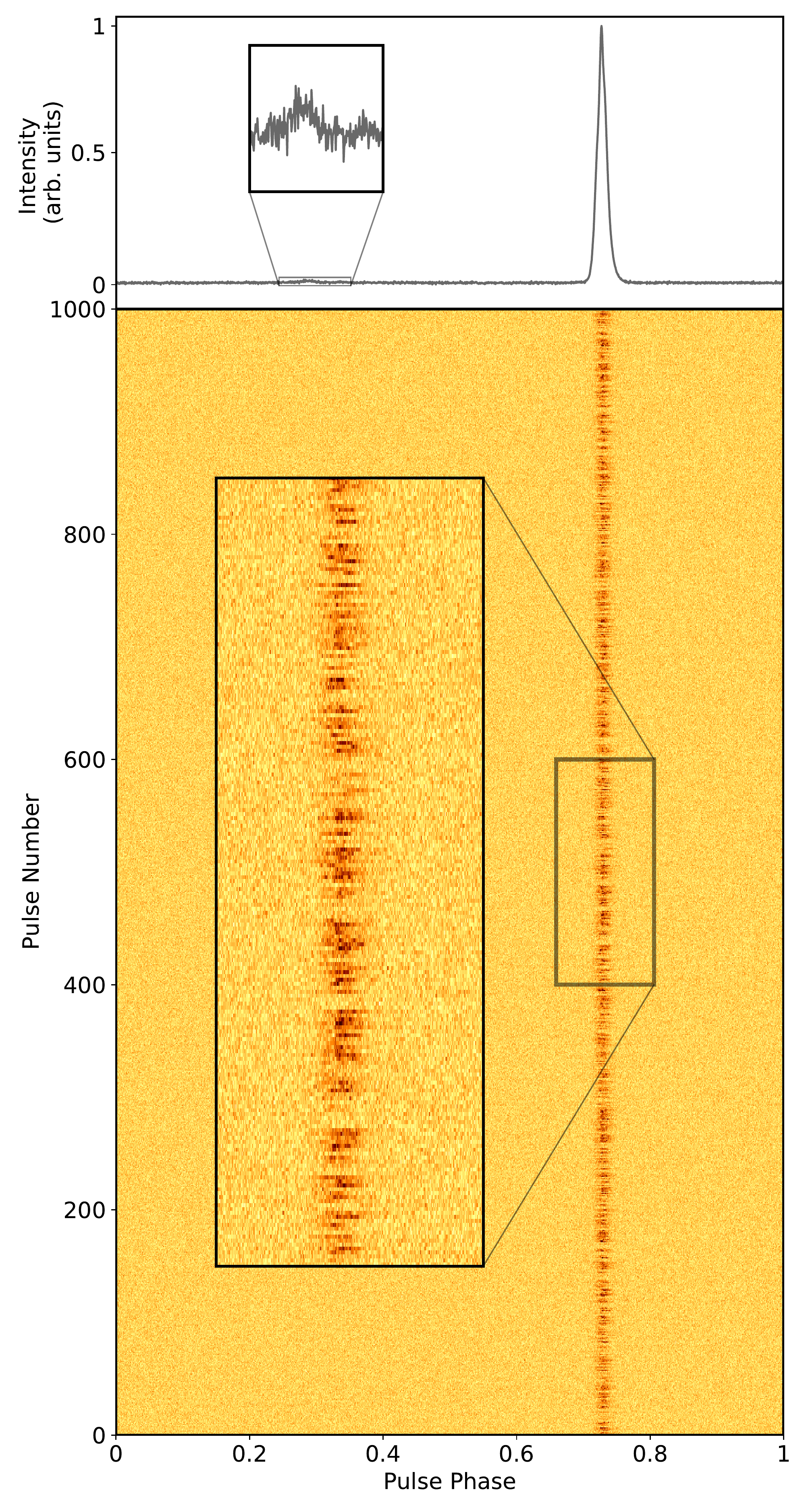}
    \caption[]{(Top) Average pulse profile PSR J1909$-$3744. The inset shows the very weak interpulse possessed by the pulsar. (Bottom) Intensities of 1000 consecutive individual pulses of PSR J1909$-$3744. The inset shows a subset of 200 of these pulses, in which clear fluctuations in pulse intensity can be observed.}
    \label{fig: 1000_SP}
\end{figure}

\section{Observations}
\label{sec: Observations}
The MeerKAT telescope in South Africa is one of four precursors to the next-generation  Square Kilometer Array (SKA) telescope. One of the major projects currently being undertaken at the site is the MeerTime project \citep{2016mks..confE..11B,2020PASA...37...28B}, which aims to take advantage of the unique combination of location and next-generation radio-telescope technology for pulsar timing science. The MeerTime project has four major science themes, one of which is the MeerTime Pulsar Timing Array (MPTA).  The aim of this theme is to search for and detect the nanohertz-frequency gravitational waves through the precision timing of MSPs.

To test the capability of a new baseband recording capability, the millicond pulsar PSR J1909$-$3744 was observed for $157$ seconds using the MeerKAT L-band receiver, which possesses a centre frequency of $1283.58$ MHz and a bandwidth of $856$ MHz, spanning $856-1712$ MHz. Fortuitously during this observation the pulse exhibited a strong scintillation event, and it was possible to post-process the data to study every single pulse with reasonable signal-to-noise ratios (as discussed below). Because of the short spin period ($\mathrm{P}=2.95\ \mathrm{ms}$) of the pulsar the data set comprised $\sim53 000$ individual pulses. 
The \textsc{ptuse} pulsar timing instrument is described in detail in \citep{2020PASA...37...28B} but it consists of four servers, each of which uses four 2TB solid state drives in parallel to sustain writing 8-bit baseband data to disk for about 40 minutes of the entire 856 MHz of dual-polarisation data with a
write speed of 3.46 GB\,s$^{-1}$. These data have been channelised by the MeerKAT F-engine using a polyphase filterbank to produce 1024 channels. 
The data was folded at the pulse period using \textsc{dspsr} \citep{2011PASA...28....1V}, and output at a resolution of $2048$ phase bins, and $1024$ frequency channels with full polarisation information, as a \textsc{psrfits} archive. The radio-frequency interference (RFI) that was present was excised using the \textsc{paz} utility in \textsc{psrchive} \citep{2004PASA...21..302H}, using a median smoothed difference algorithm. Following this, the archives were averaged in frequency to a resolution of $32$ channels in order to reduce data volume.

The archives were calibrated in polarisation through the use of Jones matrices to correct for instrumental leakage, applied through the \textsc{pac} utility of \textsc{psrchive} \citep{2010PASA...27..104V}. Further, a right-handed circular polarisation convention was adopted, and the parameters were converted from coherence, a description of polarisation through the intensity of the two linear components and their cross terms, to the more familiar Stokes parameters.

\section{Analysis \& Evidence for Mode Changing}
\label{sec: mode_changing}
To analyse the data we first investigated the pulse intensities, including modelling the pulse energy distribution and understanding the shapes of the individual pulses. Following this, we classified the pulse distribution and undertook a timing analysis. As noted above, the data were reduced to $32$ sub-bands spanning the MeerKAT L-band receiver bandwidth. Diffractive scintillation causes the pulse intensity of low dispersion measure pulsars such as PSR J1909$-$3744 to vary with frequency, and as a result the 9 sub-bands from 963 to 1204 MHz were the brightest. We focus on this frequency range for much of this analysis as it maximises the signal to noise ratios.

We first measured the signal to noise ratio of the pulses (S/N) over a pulse window of $100$ bins (corresponding to a duty cycle of $0.049$), defined to be
\begin{equation}
\label{eq: snr}
    \mathrm{S/N} = \frac{\mathrm{S} - (\mathrm{N} \cdot \mu)}{\sigma_{\rm{n}} \cdot \sqrt{\mathrm{N}}},
\end{equation}
where $S$ is the integrated intensity (fluence) in these bins, and $\mu$ and $\sigma$ are the off-pulse mean and standard deviation. This statistic was chosen as it accommodates for shape variations between pulses and other statistics can give biased measurements of S/N for weak pulses. 

If the pulse intensity distribution is Gaussian, the analytic form for the convolved distribution is
\begin{equation}
\label{eq: PDF_gauss}
     \rho_s(x)  =  \frac{1}{\sqrt{2 \pi (\sigma_{\rm{n}}^2 + \sigma^2)}} \exp \left( -\frac{1}{2}\frac{(x-\mu)^2}{ (\sigma_{\rm{n}}^2+ \sigma^2) } \right).
\end{equation}
where $\sigma_{\rm{n}}$ is the standard deviation of the noise in the data.
If a pulsar is mode changing, we can model the pulse intensities to have a bimodal distribution, in the case that both modes are Gaussian, the intrinsic intensity distribution is
\begin{align}
\label{eq: PDF_bimodal}
\rho_s(x) &= f \frac{1}{\sqrt{2 \pi (\sigma_{\rm{n}}^2 + \sigma_1^2)}} \exp \left( -\frac{1}{2}\frac{(x-\mu_1)^2}{ (\sigma_{\rm{n}}^2+ \sigma_1^2) } \right)  \nonumber \\ 
&+ (1-f) \frac{1}{\sqrt{2 \pi (\sigma_{\rm{n}}^2 + \sigma_2^2)}} \exp \left( -\frac{1}{2}\frac{(x-\mu_2)^2}{ (\sigma_{\rm{n}}^2+ \sigma_2^2) } \right),
\end{align}
where $f$ parametrizes the relative fraction of time the pulse emits in each mode. 

The distribution of the statistic in the presence of the pulsed emission will then be the convolution of pulse intensity distribution $\rho_p(x)$ and the noise distribution $\rho_n(x)$.  As the intrinsic pulse energies must be positive, the deconvolved S/N must also be positive.
\begin{equation}
\label{eq: convolution_example}
\rho_s(x)  =  \int_0^\infty dx^\prime \rho_n(x-x^\prime) \rho_p(x^\prime)
\end{equation}

It is likely that the pulse intensity distributions are not Gaussian. Indeed, \citet{2014MNRAS.443.1463S} found that the pulse energy distribution for PSR J1909$-$3744 had excess Kurtosis relative to a Gaussian distribution. We generalise the distribution by parameterizing the excess Kurtosis using parameters $\alpha_{1}$ and $\alpha_{2}$. In this case the distribution of the statistic is no longer analytic, and we also account for the fact that the pulse energy distribution must be positive:


\begin{align}
\label{eq: PDF_final}
    \rho_s&(x) = \frac{1}{C} \int_{0}^{\infty}  dx^\prime \frac{1}{\sqrt{2 \pi \sigma_{\rm{n}}^2}} \exp \left(-\frac{(x-x^{\prime})^2}{2\sigma_{\rm{n}}^2} \right)  \\ & \times \Bigg[f \frac{\alpha_{1}}{ 2^{1+1/\alpha_{1}} \sigma_{1} \gamma(1/\alpha_{1})} \exp \left(-\frac{1}{2} \displaystyle\left\lvert\frac{(x^\prime-\mu_{1})}{\sigma_{1}}\right\rvert^{\alpha_{1}}\right) \nonumber \\ &+ (1-f) \frac{\alpha_{2}}{ 2^{1+1/\alpha_{2}} \sigma_{2} \gamma(1/\alpha_{2})} \exp \left(-\frac{1}{2}\displaystyle\left\lvert\frac{(x^\prime-\mu_{2})}{\sigma_{2}}\right\rvert^{\alpha_{2}}\right)\Bigg], \nonumber
\end{align}

where $C$ is the normalising constant that ensures $\rho_p$ is normalised
\begin{align}
\label{eq: normalize_final}
    C =& \int_{0}^{\infty}  dx   \Bigg[f \frac{\alpha_{1}}{ 2^{1+1/\alpha_{1}} \sigma_{1} \gamma(1/\alpha_{1})} \exp \left(-\frac{1}{2} \displaystyle\left\lvert\frac{(x-\mu_{1})}{\sigma_{1}}\right\rvert^{\alpha_{1}}\right)  \\ &+ (1-f) \frac{\alpha_{2}}{ 2^{1+1/\alpha_{2}} \sigma_{2} \gamma(1/\alpha_{2})} \exp \left(-\frac{1}{2}\displaystyle\left\lvert\frac{(x-\mu_{2})}{\sigma_{2}}\right\rvert^{\alpha_{2}}\right)\Bigg]. \nonumber
\end{align}

Equations \ref{eq: PDF_final} and \ref{eq: normalize_final} can be used to model pulse intensities in the case they show a bimodal distribution, as would be expected for nulling or mode-changing behaviour.

In addition to analysing the pulse intensities, we analysed the pulse arrival times of weak and strong pulses at a variety of integration times. These arrival times were collected using the \textsc{pat} command of \textsc{psrchive} by cross-correlating a template against the single pulses. A high quality standard template used for the timing analysis was created from the full set of the single pulse data, another was created from the higher S/N mode of the single pulses, and a third was created from the lower S/N mode that we discuss later. All standard templates were wavelet smoothed using the \textsc{psrsmooth} tool from \textsc{psrchive} prior to timing. As the low S/N mode had, by definition, low S/N, it was difficult to make a robust template for it. Furthermore, time tagging algorithms can be biased when the S/N is low.

\begin{figure}
    \centering
    \includegraphics[width=\columnwidth]{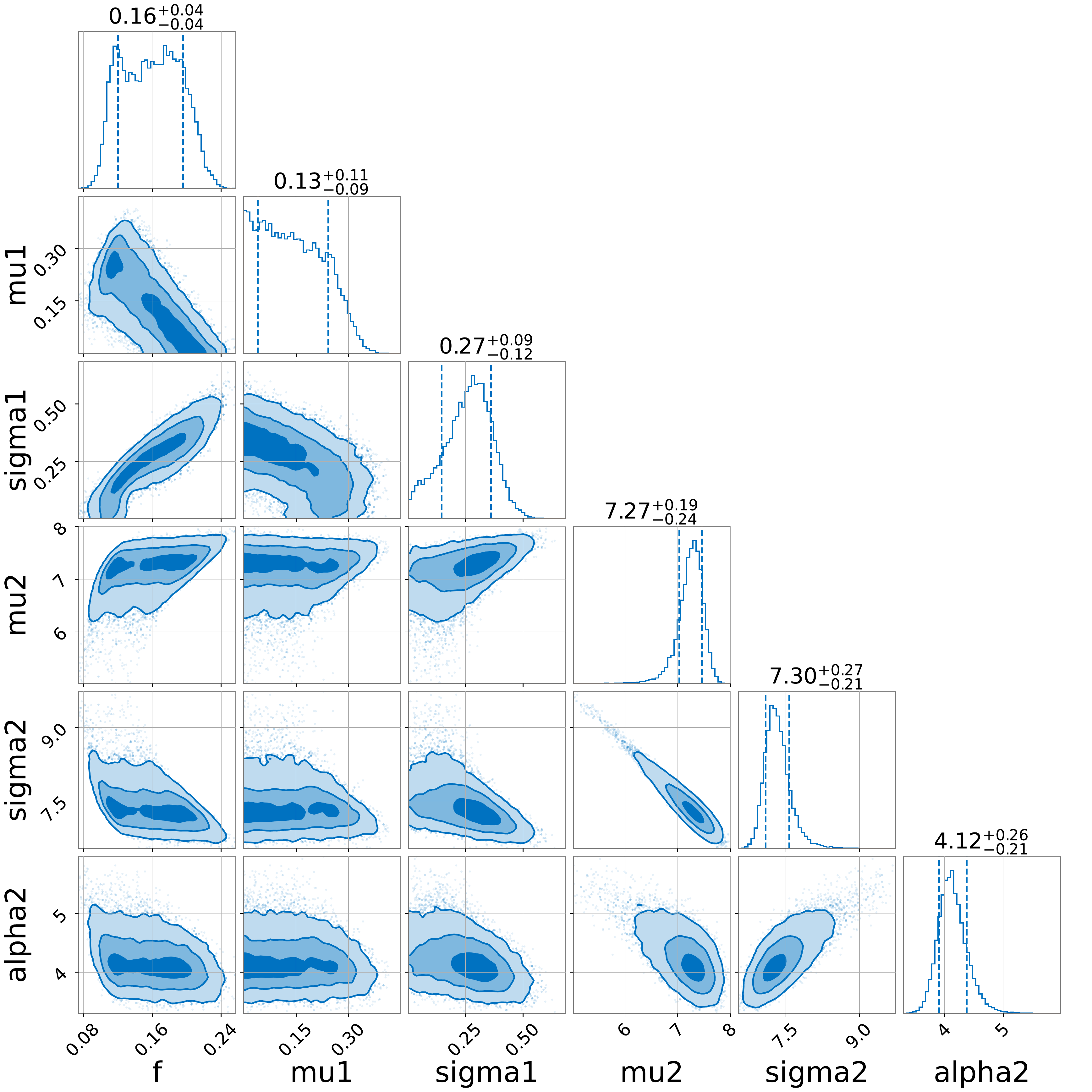}
    \caption[]{Marginalised posterior distribution for the pulse amplitude distributions of PSR J1909$-$3744. The dark and light shadings represent the $1\sigma$, $2\sigma$, $3\sigma$ intervals.}
    \label{fig: corner_plot}
\end{figure}

The clear fluctuations in flux (Figure \ref{fig: 1000_SP}) suggest bimodal distribution of pulse intensity, consistent with what would be expected of a mode-changing pulsar. We refer to these modes as the weak and strong modes. We used Bayesian parameter estimation, using the \textsc{Bilby}\footnote{\url{https://git.ligo.org/lscsoft/bilby}} package \citep{Ashton+19}, to characterise the properties of the intensity distribution described in Equation \ref{eq: PDF_final}, with six parameters representing the mode-changing factor $f$, the mean S/N of the first and second modes, $\mu_{1}$ and $\mu_{2}$, the standard deviations of the modes, $\sigma_{1}$ and $\sigma_{2}$, and an exponent that models the kurtosis strong component, $\alpha$.
We find strong evidence for pulse moding in single pulse observations of PSR J1909$-$3744 (the log Bayes factor that strongly supports bimodality over a single mode is: $\rm{\ln}\mathcal{B}=23.51$) through variations in flux, with the relative intensities creating a distribution ranging from S/N values of $-3.78$ to $24.65$, using the S/N defined in Equation \ref{eq: snr}.

\begin{figure}
    \centering
    \includegraphics[width=\columnwidth]{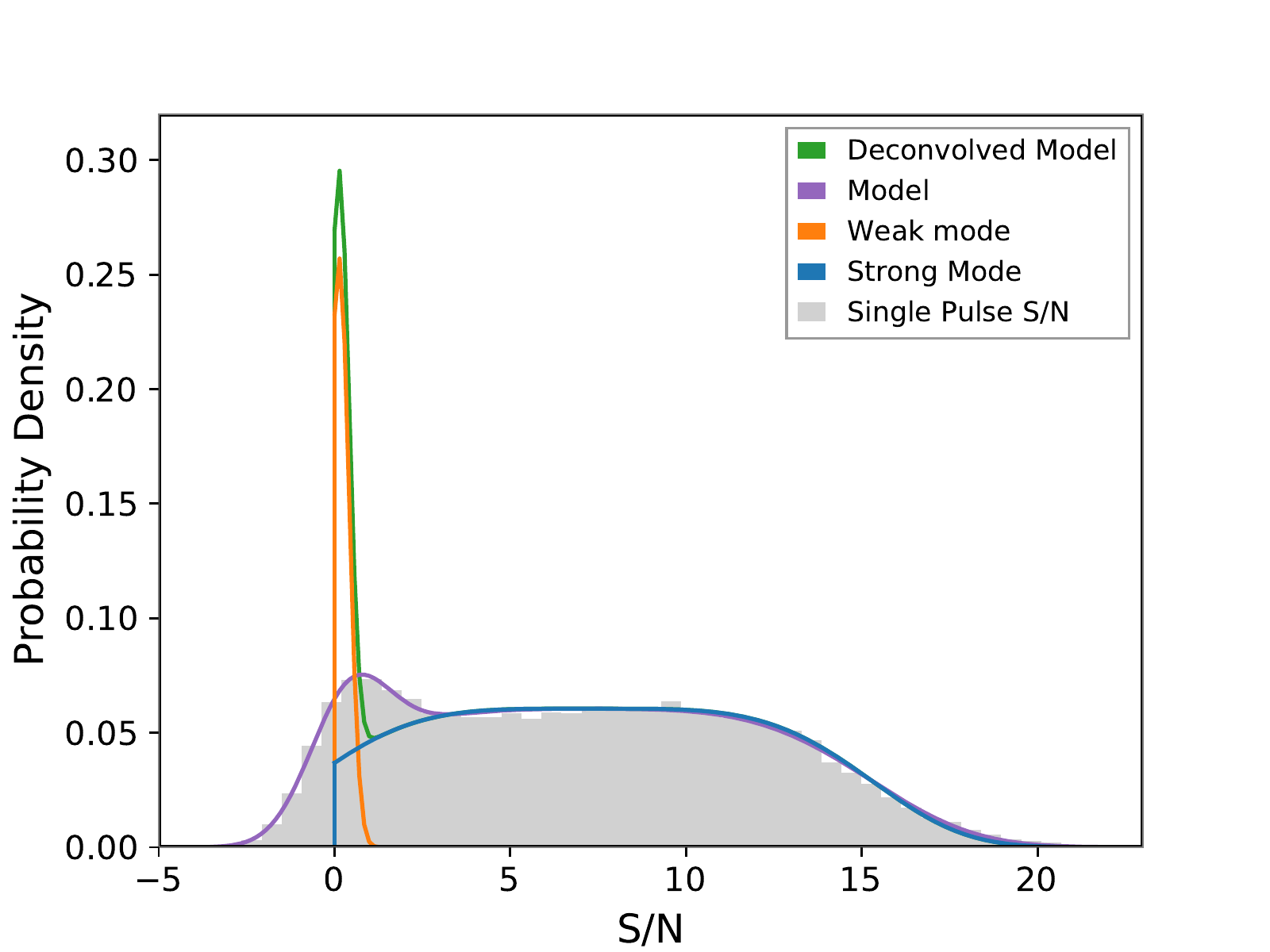}
    \caption[]{Fits to the probability density histogram of S/N, including the maximum likelihood a posteriori fit (purple), the same fit deconvolved from the observed noise (green), and the strong (blue) and weak (orange) emission modes.}
    \label{fig: snr_overlay}
\end{figure}

The marginalised posterior distributions of this parameter estimation can be found in Figure \ref{fig: corner_plot}.  Figure \ref{fig: snr_overlay} presents the S/N density histogram overlaid with the probability distribution function (PDF) and the two modes deconvolved from a noise Gaussian with parameters $\mu_{\mathrm{n}} = 0$ and $\sigma_{\mathrm{n}} = 1.1$. It is expected that the S/N of the noise would present a Gaussian distribution of zero mean and unit variance, however, in an observation of the off pulse region for the single pulses a standard deviation of $\sigma_{\mathrm{n}} = 1.1$ presented a better fit.

An additional investigation into the relative strength of the small interpulse \citep{2003ApJ...599L..99J} present in the pulse emission showed a marginal but statistically significant difference in strength dependent on which mode it was observed in. The mean fluence, $\phi$, of the interpulse in each mode was found to be $\phi_{\rm{IP,strong}} = 3.94 \pm 0.02$ and $\phi_{\rm{IP,weak}} = 4.19 \pm 0.05$.

\begin{figure}
    \centering
    \includegraphics[width=\columnwidth]{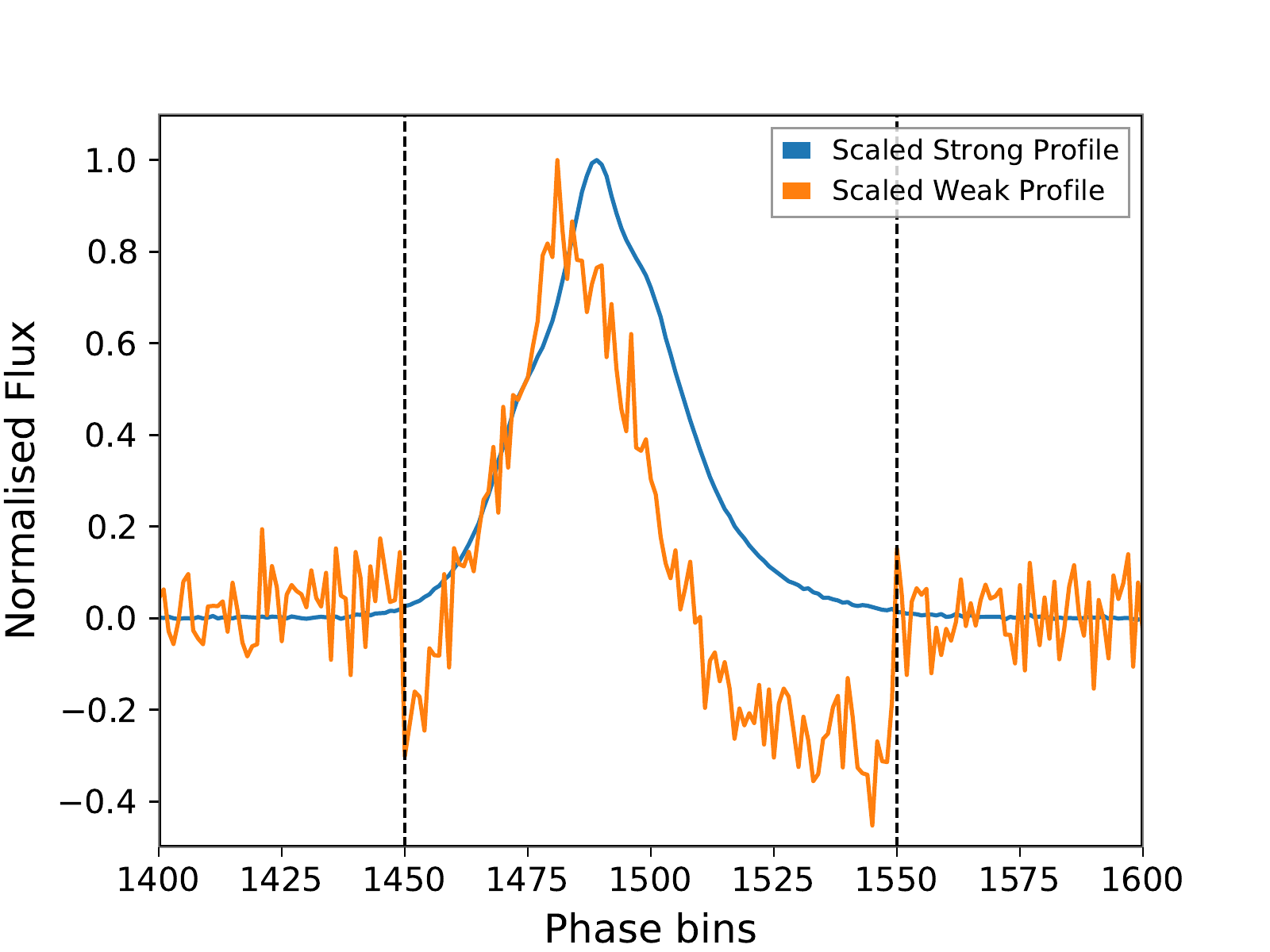}
    \caption[]{Scaled pulse profiles of the weak (orange) and strong (blue) modes. The profiles are formed from the sum of the catalogues of each mode, with their respective baselines subtracted. The peaks of the modes appearing in distinct phase bins is an indication that the pulses from the weak mode are arriving earlier than those found in the strong mode. The negative values seen in the weak mode are artefacts of the S/N algorithm selection, corresponding to the pulse window selection marked by the dashed lines.}
    \label{fig: pulse_profiles}
\end{figure}

In order to study the morphology of the two modes we produced integrated pulse profiles, formed from pulses with S/N $\geq 1.375$ and S/N $< 1.375$, corresponding to a value where the probability of a weak mode pulse existing in the strong mode distribution is marginal ($<0.01\%$). Following this condition, the strong mode was found to dominate the majority of the sample, with $44056$ of the single pulses included for analysis, compared to $8944$ in the weak mode. A comparison of the scaled, summed pulse profiles (Figure \ref{fig: pulse_profiles}) indicated that the weak mode is present at earlier in phase than the strong mode, calculated to be at an average separation of $9.26\, \pm 3.94\, \mu \rm{s}$ from the strong mode.

\begin{figure}
    \centering
    \includegraphics[width=\columnwidth]{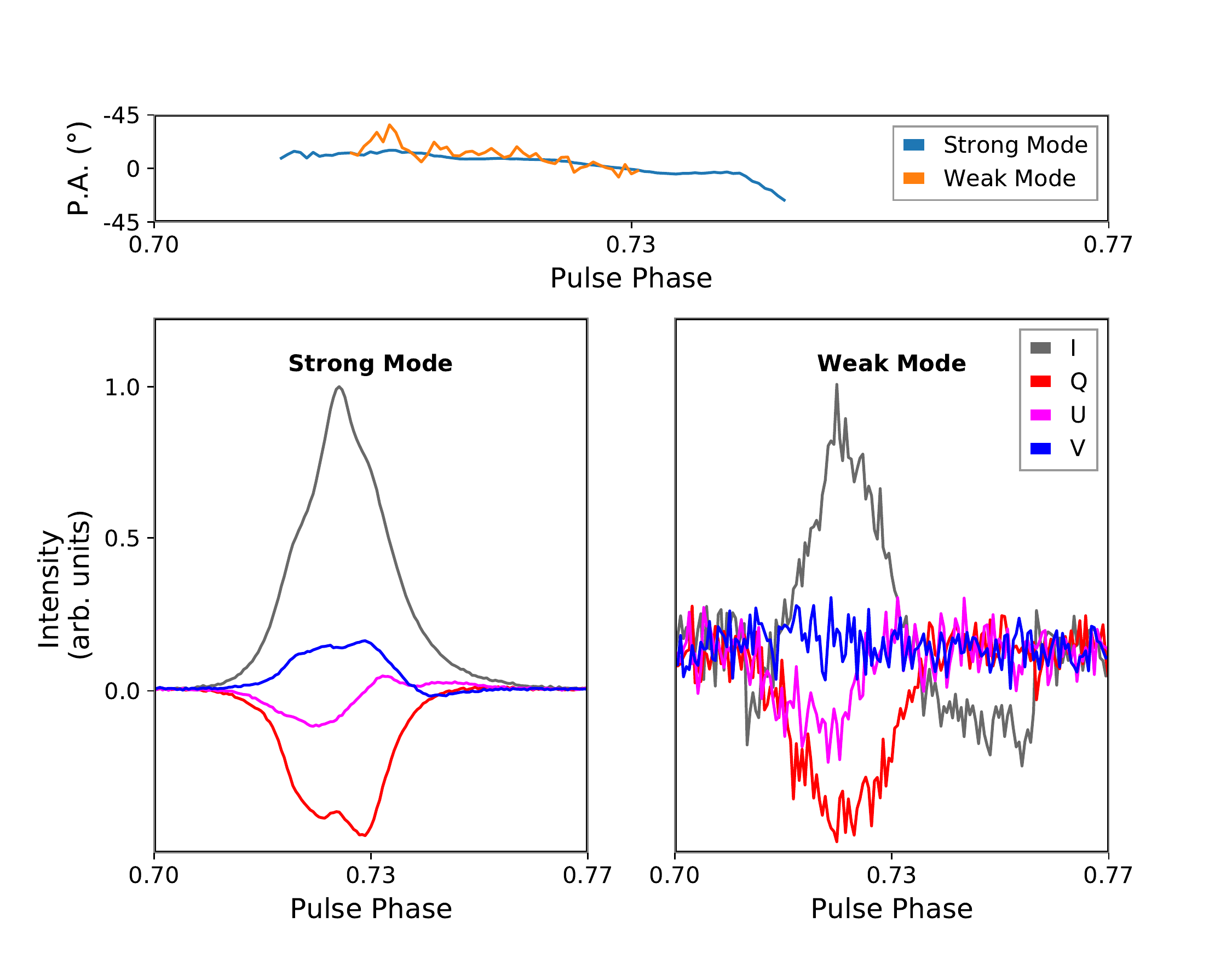}
    \caption[]{Position angle (top) and integrated polarimetric profiles (bottom) for the strong (left) and weak (right) modes. The weak mode clearly shows a lack of circular polarisation (Stokes V, blue), as well as a missing secondary peak in Stokes Q (red). The negative values seen in the weak mode Stokes I profile are artefacts of the S/N algorithm.}
    \label{fig: StokesModes}
\end{figure}

Previous studies of mode changing, including the study of the previously known mode-changing MSPs, have identified modulations both in the total intensity and the polarisation components. The polarisation parameters (Figure \ref{fig: StokesModes}) show clear differences that cannot be completely attributed to differences in S/N. The most distinct of these is the almost complete lack of circular polarisation and the absence of the second peak in the Stokes Q profile of the weak mode, as well as the difference in relative flux density of the Stokes Q and U profiles between the modes.

An investigation into the periodicity of the pulse strength fluctuation showed evidence for a frequency, $\rm{f}=0.042$ per pulse period, at which the emission mode changed. A Lomb-Scargle periodogram revealed that the single pulse distribution presented a periodic signal corresponding to $24.07$ individual pulses. Phase-folding the data at this frequency revealed an approximately sinusoidal signal confirming that the detected periodicity was present in the data, and a false-alarm probability (FAP) was calculated to be $\rm{FAP}=0.006$. We note, however, that the pulse intensity distribution is not deterministic, and the noise is non-Gaussian, which is what the FAP assumes.

\section{Jitter and its Variation}
\label{sec: Jitter Variation}
Pulse-to-pulse morphology changes, otherwise known as jitter, are widely studied in pulsars used for PTAs as they provide a limit to the timing precision that is achievable for the pulsar \citep{2011MNRAS.418.1258O, 2012MNRAS.420..361L,2012ApJ...761...64S, 2014MNRAS.443.1463S, 2019ApJ...872..193L, 2021MNRAS.502..407P}. As one of the most precisely timed and brightest pulsars, PSR J1909$-$3744 has naturally been included in these studies. The most recent study \citep{2021MNRAS.502..407P} into the jitter properties of this pulsar was performed with the same observing system as this study, and presented a jitter-limit of $9 \pm 3$ ns. 

We follow the method presented in this work to model the jitter properties of the pulsar, where the jitter noise ($\sigma_{\rm{J}}$) is expressed as the quadrature difference of the RMS of observed TOAs and an idealised set of simulated TOAs created from the \textsc{libstempo} python package expressed in:
\begin{equation}
    \label{eq: quad_diff}
    \sigma_{\rm{J}}^{2}(\rm{T}_{\rm{sub}}) = \sigma_{\rm{obs}}^{2}(\rm{T}_{\rm{sub}}) - \sigma_{\rm{sim}}^{2}(\rm{T}_{\rm{sub}});
\end{equation}
where $\rm{T}_{\rm{sub}}$ is the sub-integration time of the observation in seconds, $\sigma^{2}_{\rm{obs}}$ is the variance of the observed arrival times, and $\sigma^{2}_{\rm{sim}}$ is the variance of the simulated arrival times. The simulated arrival times are found by creating a set of idealised TOAs that match the model. Following this, uncertainties are added to the idealised set of TOAs with gaussian noise, consistent with the magnitude of the uncertainties in the observed TOAs. As conducted in \citet{2021MNRAS.502..407P}, we created $1000$ realisations of these simulated data sets, and the mean $\sigma_{\rm{sim}}^{2}(\rm{T}_{\rm{sub}})$ is taken to derive the quadrature difference. To compare to other studies, we scale our value to what would be expected in a $1\,\rm{hr}$ observation:
\begin{equation}
    \label{eq: jitter_equation}
    \sigma_{\rm{J}}(1\rm{hour}) = \sigma_{\rm{J}}(\rm{T}_{\rm{sub}})/\sqrt{3600/(\rm{T}_{\rm{sub}})}.
\end{equation}

The corresponding uncertainty to this value can be evaluated as the quadrature sum of the sample error ($\sigma_{\rm{ToA}}$) and the error associated with the radiometer noise ($\sigma_{\rm{rad}}$), such that
\begin{equation}
    \label{eq: jitter_error}
    \rm{err}_{\sigma_{J}} = \sqrt{\sigma_{\rm{ToA}}^{2}+\sigma_{\rm{rad}}^{2}}.
\end{equation}

We find evidence for different levels of jitter noise when examining the strong mode ($7.29 \pm 0.10$ ns), and the total sample of single pulses ($8.20 \pm 0.14$ ns), recorded at the smallest subintegration length where the pulse could be confidently detected for each sample. While both fall within the large uncertainty given by the previous work, the strong mode does not fall in the range of the two next most recent studies, $8.6 \pm 0.8$ ns \citep{2014MNRAS.443.1463S} and $14 \pm 0.5$ ns \citep{2019ApJ...872..193L}.

\begin{figure}
    \centering
    \includegraphics[width=\columnwidth]{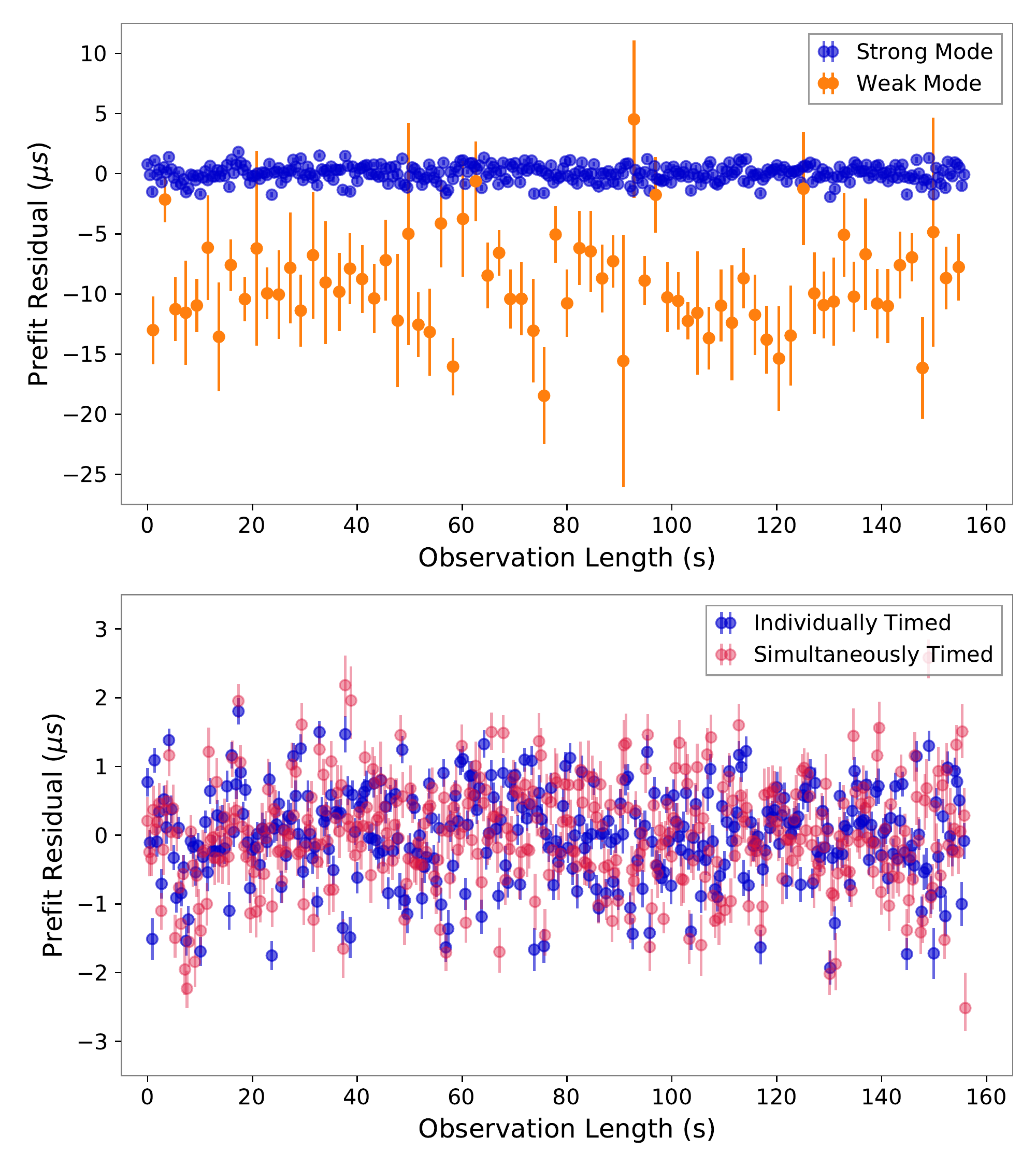}
    \caption[]{Top panel: Timing residuals calculated  for the strong (blue) and weak (orange) modes. Each data point represents the average of $128$ consecutive pulses of that mode. The remaining pulses that were unable to represent a full $128$ pulse sample were not included. The negative mean residual for the weak mode is clear evidence that it is arriving earlier. 
    Bottom panel: Timing residuals for the strong mode determined by separating the strong pulses (blue), and by timing simultaneously with the weak mode in folded data (red). The drifting phenomena that can be seen in the residuals has been noted before in pulsars including PSRs J1909$-$3744 and J1713$+$0747 \citep[e.g.][]{2014MNRAS.443.1463S}.}
    \label{fig: mode_residuals}
\end{figure}

As expected, there is evidence to suggest that a portion of the jitter noise is due to the multiple modes. Standard pulse profiles are created from high-quality observations, representing the average pulse profile over those observation times. Should multiple emission modes contribute to the construction of such a standard, timing uncertainties will emerge where the emission modes are timed as individual samples, as is apparent in Figure \ref{fig: mode_residuals}. To confirm this, three timing samples were created with the strong mode, weak mode, and total sample. Each sample was fully averaged in frequency and measured in total intensity. We then examined how the timing precision improved with an increasing number of pulses averaged in each sub-integration, and found that the strong mode consistently showed a $\sim10\%$ improvement compared to the total pulse regime at all degrees of time averaging. The weak mode possessed a consistently higher timing uncertainty than the other two regimes, this is thought to be a combination of both the low S/N present in the sample, and the biases in measuring arrival times discussed above.
Figure \ref{fig: jitter_comps} further illustrates this through measurements of $\sigma_{\rm{J}}(1\rm{hour})$ for the full sample and the strong mode. The consistently lower jitter values emerging in the strong sample through degrees of time averaging confirm the difference in the values are due to the convolution of multiple emission modes, and not only the removal of low S/N pulses. 

The $\sim23\%$ strong mode reduction and $\sim10\%$ total sample reduction in jitter noise as compared to the previous reported value is of interest as they may be an indication of possible ways that the current limits of timing precision for some pulsars can be leveraged or improved.

\subsection{Simultaneous timing of two modes in folded data}
\label{sec: Simultaneous}
Here we develop a new method for accounting for multiple modes when timing a pulsar. This method is applicable to standard pulsar timing observations, in which the data are folded at the pulse period of the pulsar and many pulses are averaged.

\begin{figure}
    \centering
    \includegraphics[width=\columnwidth]{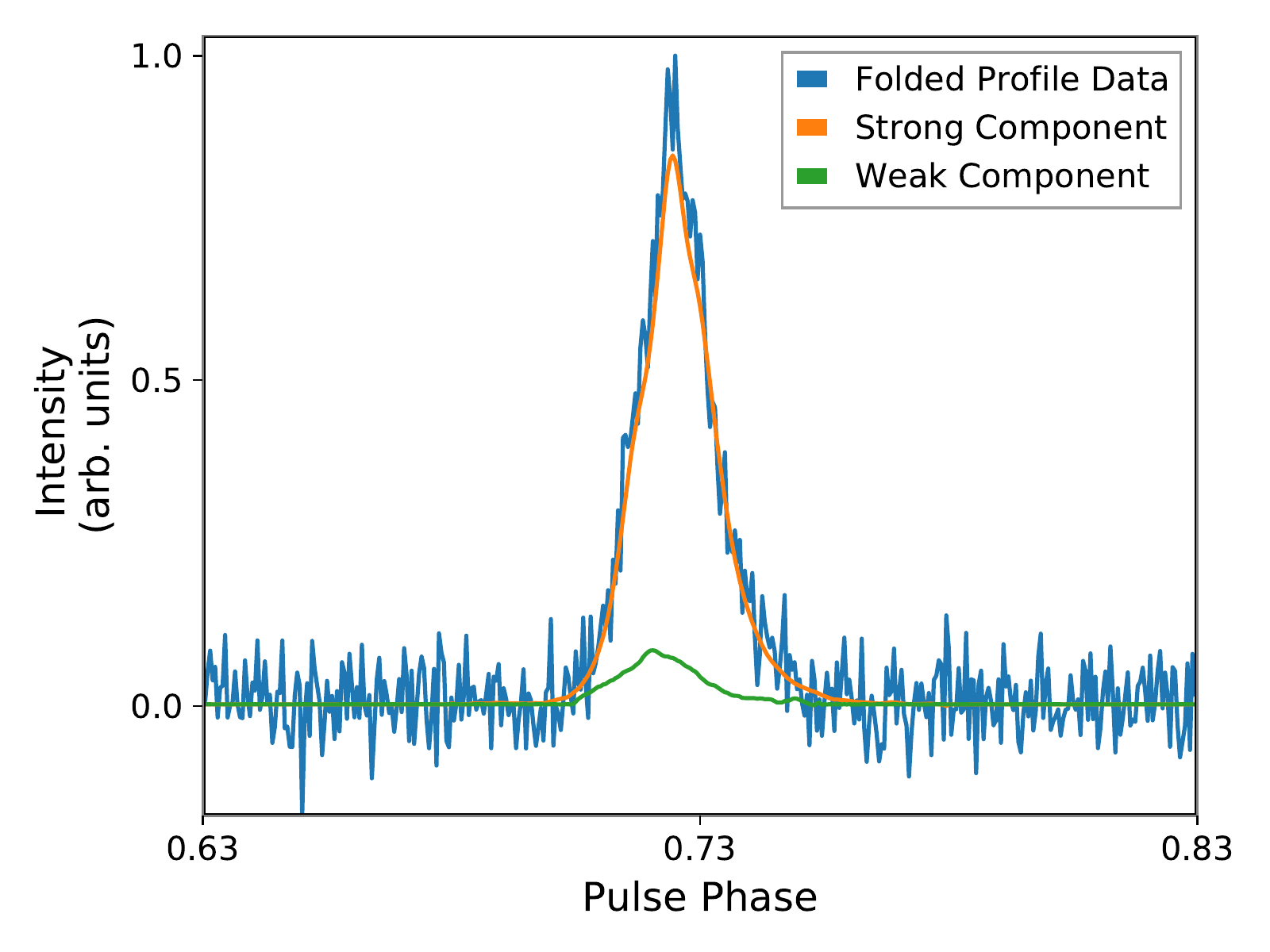}
    \caption[]{The measured model contributions of the weak mode (green) and the strong mode (orange) to the average pulse profile data (blue) in a folded sub-integration. Using this method, ToAs of the weak and strong modes are be extracted simultaneously and show similar timing improvements to isolating the modes with baseband data.}
    \label{fig: TwoMode_Temps}
\end{figure}

We created templates for the two modes by averaging the full sample of their individual pulses, smoothing the on-pulse regions using a Savitzky–Golay filter, and setting the off-pulse regions to zero. We form a model of the average pulse in an observation by allowing these templates to vary independently in phase and amplitude, before summing them, as shown in Figure \ref{fig: TwoMode_Temps}. These amplitudes and phase offsets for the two modes were measured for each sub-integration using \textsc{Bilby} \citep{Ashton+19}. The resulting measurements of phase offsets for each mode are then converted into ToAs.

Uniform prior probability distributions were used for the phase and amplitude of the strong mode, while Gaussian priors were used to describe the amplitude and phase of the weak mode relative to the strong mode. These Gaussian priors were determined from our analysis of the modes separately. The prior for the phase offset of the weak mode relative to the strong mode, was determined from the mean and standard deviation of the offset in timing residuals in Figure \ref{fig: mode_residuals}.

The resulting timing residuals from this method are consistent on average with those measured by timing the modes independently, as shown in the lower panel of Figure \ref{fig: mode_residuals}. However, by timing the strong mode independently using 128 pulses, the ToAs do not correspond directly to consecutive sub-integrations. Modelling the folded data in this way results in a jitter noise value for the strong mode ToAs of $\sigma_{\rm{J}} = 6.66 \pm 0.23$\,ns. To assess the significance of this result, more familiar timing methods (\textsc{pat}/\textsc{tempo2}) were used on a dataset of the same subintegration length, made of strong mode pulses and timed with a strong mode template, as well as a dataset of the total sample timed with a template made from the average representative pulse of the weak and strong modes. From this, a comparable level of jitter noise was found from the total sample, $\sigma_{\rm{J}} = 6.73 \pm 0.23$\,ns, and a marginal reduction in jitter noise was found from timing the strong mode with a strong mode template, $\sigma_{\rm{J}} = 6.16 \pm 0.24$\,ns. However, taking into account the need to establish an effective jitter noise for partial datasets, as is discussed in Section \ref{sec: Discussion}, the strong mode jitter noise is more accurately expressed as $\sigma_{\rm{J,eff}} = 6.76 \pm 0.24$\,ns, a similarly comparable jitter value.


\section{Discussion}
\label{sec: Discussion}
Mode changing and nulling are thought to be different manifestations of the same phenomenon, linked to a global change of the pulsars magnetosphere. Until recently neither had  been detected in the millisecond pulsar regime. This changed with the discovery of the multi-modal nature of B1957$+$20 \citep{2018ApJ...867L...2M}, a $1.6$ ms pulsar. The emission characteristics discovered in PSR J1909$-$3744 provide further evidence that mode changing and nulling, usually found in slower pulsars, extends into the millisecond regime and on far smaller timescales.
In contrast to B1957$+$20, we find that the weak mode is present earlier in pulse phase than the strong mode. 

The variation we observe between modes supports a commonly accepted mechanism for the pulsar mode changing phenomena.The polarimetric profiles of both modes exhibit similarly flat position angles with slight negative trends, however there exists a distinct lack of circular polarisation in the weak mode. The polarimetric parameters also suggest a relative difference in the ratio of the linear polarisation components across the pulse, the weak mode possessing a mean ratio of U and Q ($\Delta$) $3.6$ times greater than the strong mode ($\Delta_{\rm{Strong}}=0.14; \Delta_{\rm{Weak}}=0.50$). That the changes we observe are present in both the total intensity and the polarisation components lends evidence to the theory that mode changing activity stems from a global reconfiguration of the pulsar magnetosphere, either in the form of a changing polar cap geometry, or a varying particle density distribution as suggested by \citet{2010MNRAS.408L..41T}. As the mode changing occurs without any  intermediate emission, we are able to infer that the mechanism behind this reconfiguration must occur at a timescale no larger than the time separation between the observation of a strong mode pulse and a subsequent weak mode pulse, allowing us to place an upper limit on the magnetosphere reconfiguration time of $\tau = 2.941$ ms.

\begin{figure}
    \centering
    \includegraphics[width=\columnwidth]{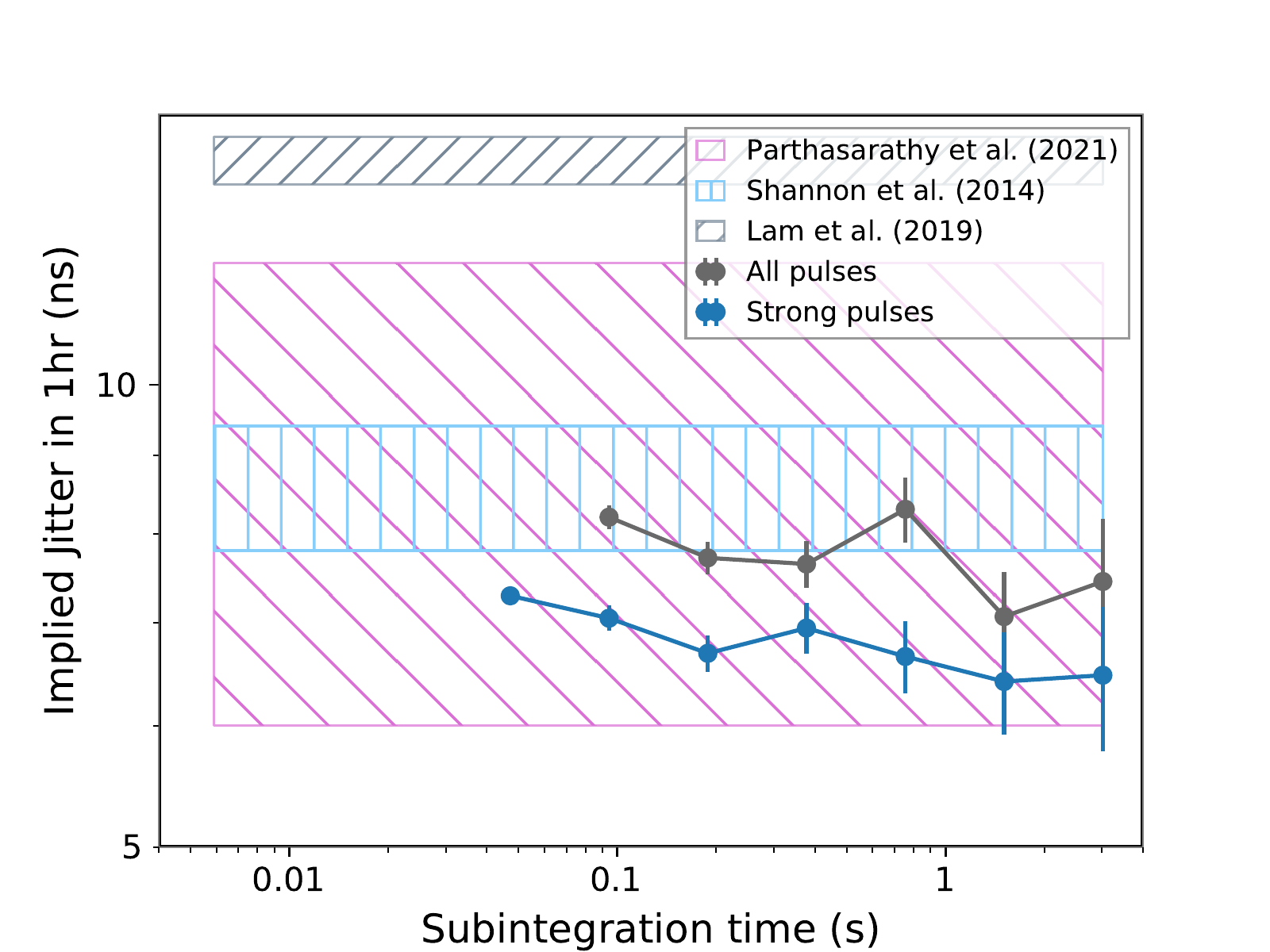}
    \caption[]{Jitter measurements as a function of the sub-integration time for the entire single pulse sample (grey) and the strong mode sample (blue), as compared to the three previously obtained jitter measurements.}
    \label{fig: jitter_comps}
\end{figure}

A standard template of this pulsar will be representative of the convolved average of the two modes, separating these modes and timing with their own representative standards results in a more precise timing solution, but at the expense of the flux that the unused mode provides. The timing analysis performed in this study shows that in the isolation of the strong mode, a moderate improvement in timing precision of $\sim 10 \%$ can be achieved through the cross correlation of the strong pulses against a truly representative standard template. Supporting this are the results of the jitter analysis, with Figure \ref{fig: jitter_comps} demonstrating that the jitter noise present in the sample lacking the convolution of modes is consistently lower ($7.29 \pm 0.10$ ns) than the full sample ($8.20 \pm 0.14$ ns). The jitter noise of the full sample is shown to be consistent with two of the three previously reported values for this pulsar, whereas the strong sample is only consistent with a single previous study. This inconsistency is expected for the strong sample, as the mode-changing characteristic would have been convolved with the jitter noise in previous studies. 
As the expected timing uncertainty will decrease proportionally with the square root of the number of pulses in the timing sample, a caveat must be included in reporting the jitter noise for the strong mode, such that there also exists an effective jitter noise:
\begin{equation}
    \label{eq: eff_jitter}
    \sigma_{\rm{J,eff}} = \sigma_{\rm{J}} \times \left( \frac{N_{\rm{selected}}}{N_{\rm{all}}} \right)^{-0.5}.
\end{equation}

This effective jitter noise is what is expected in practice over multiple observations, due to the loss in flux that comes from down-selecting the full sample to just include the strong mode, as opposed to the theoretical jitter noise reported in Section \ref{sec: Jitter Variation}. We find here that the effective jitter noise still shows some improvement, but not to as significant a degree, $\sigma_{\rm{J,eff}} = 8.00$ ns.

However, simultaneously timing both modes in folded data allows us to account for the added jitter noise in the average pulse profile that is attributed to the presence of the weak mode. Timing the modes individually by separating pulses with a threshold signal-to-noise ratio, may be erroneously classifying some weak mode pulses as strong mode. Simultaneous timing of the two modes in folded data, as described in Section \ref{sec: Simultaneous}, may help to mitigate this and results in a jitter noise of $\sigma_{\rm{J}} = 6.66 \pm 0.23$ ns, which is comparable to timing the strong mode individually, which at the same subintegration length, achieves an effective jitter noise value of $\sigma_{\rm{J,eff}} = 6.76 \pm 0.24$\,ns. The implications of extracting simultaneous ToAs of the two modes, are that timing advantages for mode-changing pulsars are able to be capitalised on with regular timing observations. A relatively small amount of baseband recorded data is required only to identify the two modes, create representative templates, and determine priors for the amplitude and phase of the weaker mode relative to the stronger.

Similar campaigns to the one undertaken in this study may result in valuable insights into millisecond pulsars emission structures, and offer opportunities to gain greater timing precision. In precision timing studies of millisecond pulsars, PSR J1909$-$3744 is consistently one of the most precisely timed \citep{2020MNRAS.499.2276L, 2021ApJS..252....4A, Reardon+21}, with sub-microsecond root mean square (rms) timing residuals often achieved. This precision leads it to be one of the most important pulsars in the search for the SGWB, with publications \citep{NanoGravGWB,2021arXiv210712112G} suggesting it is the first (NANOGrav) and third (PPTA) largest contributors to the detection efforts. The primary goal of pursuing greater timing precision in PTAs is the detection of nanohertz gravitational waves and this background. The MeerTime PTA is poised to contribute significantly to this endeavour in the coming years through the regular observation of the highest number of pulsars of any current PTA, the ability of MeerKAT to observe Southern Hemisphere sources, and its sensitivity with respect to other southern hemisphere observatories. Another advantage that the MeerTime PTA is able to exploit is that of the baseband recording system on the MeerKAT telescope. By taking baseband observations of other key millisecond pulsars, we may be able to shed light on the intrinsic emission structure, which could offer opportunities to gain greater precision in pulsar timing.

Single pulse analysis of millisecond pulsars greatly increases the complexity and volume of data needed to be recorded. A typical pulsar timing analysis proceeds on mean pulse profiles usually folded for typically 8-20 seconds and binned with 1024 channels leading to output data rates of $\sim$ 1 MB/s. If the time resolution is increased to 2048 channels, and individual pulses stored, the output data rate increases by over three orders of magnitude for PSR J1909--3744. To remain practicable, it would probably be necessary to record and analyse single-pulse data in near real time and only store an average profile and some statistics and arrival times. It is also worth considering if any improvements to the arrival times O(ns) are negated by uncertainties in the pulsar dispersion measure, as these lead to uncertainties in the infinite frequency pulse arrival times used in timing analyses. Small dispersion measure pulsars such as PSR J1909--3744 rarely emit strongly over the entire MeerKAT L-band (856-1712 MHz) and this leads to higher uncertainties in the dispersion measure.

While the increase in sensitivity that a $\sim10\%$ timing improvement in PSR J1909$-$3744 would award is individually negligible, a similar increase in all MSPs in the MeerTime PTA would result in a sensitivity increase that becomes significant through the coming decade. Although, the assumption that all PTA MSPs are capable of achieving a similar timing improvement through the use of a baseband recording system is unfounded. While mode-changing phenomena has been detected now in three MSPs, there is no reason to suggest this is common among the MSP population, and this will only be confirmed as more high S/N single pulse studies are performed. Further, many MSPs are not bright enough to confidently detect single pulses, creating further issues. However, we must also recognise that the moding detected in this pulsar is not particularly strong, with only a small fraction ($f=0.15 \pm 0.01$) of the pulses present in the secondary weak mode, and the modes differ largely in intensity as opposed to distinct shape changes. Should an MSP be identified with comparable timing precision to PSR J1909$-$3744, and possess more distinct moding at a higher fraction, the potential timing improvement could be far more significant. To conservatively demonstrate this, we have selected a sub-sample of 30 jitter dominated candidates from the MeerTime PTA that are the most likely to produce results from a similar analysis, selected as they each possess an average single-pulse S/N of greater than unity.

\begin{figure}
    \centering
    \includegraphics[width=\columnwidth]{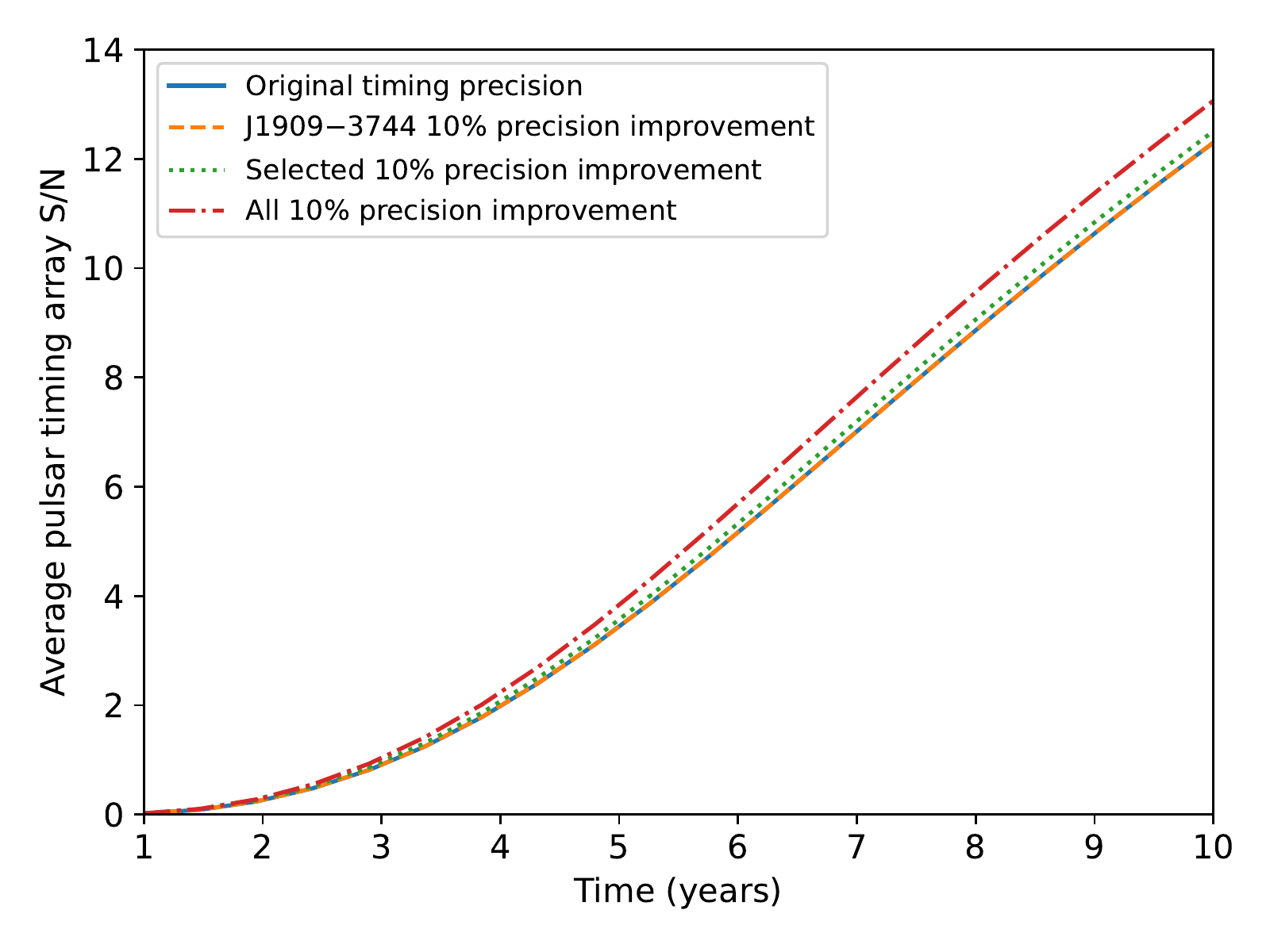}
    \caption[]{The average MeerTime PTA S/N including 89 MeerTime pulsars each observed with a cadence of $26\,{\rm yr}^{-1}$, beginning from the inception of the MeerTime program. The signal assumed has a gravitational wave background amplitude of $A=2\times10^{-15}$ and $\alpha=-2/3$, assuming a strain spectrum of $h_c(f) = A f^\alpha$. Here, the blue line represents the sensitivity of the PTA as it currently operates. The dashed orange line represents the virtually negligible improvement that would occur from including a long term baseband timing study of PSR J1909$-$3744 in the PTA. The red line represents the sensitivity if all pulsars in the PTA improve by the same margin as PSR J1909$-$3744 through a baseband timing study, and the green line represents the sensitivity increase should only the jitter dominated candidates show this same improvement.}
    \label{fig: GWB_sensitivity}
\end{figure}

The outcomes of altering the timing precision of the PTA through these methods can be seen in Figure \ref{fig: GWB_sensitivity}. The increase in sensitivity that comes as a result of the improvement of the jitter dominated candidates, while not as distinct as an improvement of the entire PTA, is certainly not negligible and should be pursued further.

\section{Conclusion}
\label{sec: Conclusion}
Through an analysis of $53000$ baseband recorded single pulses, PSR J1909$-$3744 shows strong evidence of mode-changing or nulling behaviour. Identified in this behaviour are two modes, differentiated by their relative S/N. A bayesian likelihood analysis was performed, the outcome of which strongly favored mode-changing behaviour as opposed to nulling. The lower S/N mode arrives earlier than its counterpart by $9.26\,\mu$s, leading to a broadening effect of the average pulse and increasing scatter in arrival times. In a timing analysis it was found that by timing only the high S/N mode, a consistent increase in timing precision of $\sim10\%$ could be achieved. In an analysis of the single pulse variable morphology, or jitter, we found lower values than expected. This included the finding that the jitter values of the strong mode are consistent with only one of the last three studies of this pulsar. We have demonstrated that by simultaneously timing the weak and strong modes in folded data, timing precision consistent with timing the strong mode can be extracted. This implies improvements in timing precision can be achieved in regular, folded observations of mode-changing pulsars, using only a small portion of baseband data to characterise the modes.

\section*{Acknowledgements}
MM, MB, DR, and RS acknowledge support through ARC grant CE17010004. RMS acknowledges support through ARC Future Fellowship FT190100155.
The MeerKAT telescope is operated by the South African Radio Astronomy Observatory (SARAO), which is a facility of the National Research Foundation, an agency of the Department of Science and Innovation. 
PTUSE was developed with support from the Australian SKA Office and Swinburne University of Technology, with financial contributions from the MeerTime collaboration members.
This work used the OzSTAR national facility at Swinburne University of Technology. 
OzSTAR is funded by Swinburne University of Technology and the National Collaborative Research Infrastructure Strategy (NCRIS). 

\section*{Data Availability}
The data used for this analysis will be provided on request.  

\bibliographystyle{mnras}
\bibliography{ref} 

\bsp	
\label{lastpage}
\end{document}